\documentclass[fleqn,usenatbib]{mnras}
\usepackage{newtxtext,newtxmath}
\usepackage[T1]{fontenc}

\DeclareRobustCommand{\VAN}[3]{#2}
\let\VANthebibliography\thebibliography
\def\thebibliography{\DeclareRobustCommand{\VAN}[3]{##3}\VANthebibliography}

\usepackage{graphicx}	
\usepackage{amsmath}	
\usepackage{multicol}   
\usepackage{bm}		    
\usepackage{pdflscape}	
\usepackage[T1]{fontenc}
\usepackage{ae,aecompl}
\usepackage{newtxtext,newtxmath}
\usepackage{booktabs}  

\newcommand{\kms}{km s$^{-1}$}
 
\newcommand{\metal}{\mbox{[Fe/H]}}

\newcommand{\kmsec}{\mbox{km~s$^{\rm -1}$}}

\title[A New Confirmed UMP Atari Disk star]{{A Strontium-rich Ultra Metal-poor Star in the Atari Disk Component}\thanks{This paper includes data gathered with the 6.5\,m Magellan Telescopes located at Las Campanas Observatory, Chile.}}
\author[Mohammad K.\ Mardini et al.]{
Mohammad K.\ Mardini,$^{1,2,3}$\thanks{E-mail: \href{mailto:mmardini@zu.edu.jo}{mmardini@zu.edu.jo}}
Anna Frebel,$^{2}$
Anirudh Chiti,$^{4,5}$
\\
$^{1}$Department of Physics, Zarqa University, Zarqa 13110, Jordan\\
$^{2}$Department of Physics and Kavli Institute for Astrophysics and Space Research, Massachusetts Institute of Technology, Cambridge, MA 02139, USA\\
$^{3}$Joint Institute for Nuclear Astrophysics–Center for Evolution of the Elements (JINA-CEE), East Lansing, MI 48824, USA\\
$^{4}$Department of Astronomy $\&$ Astrophysics, University of Chicago, 5640 S Ellis Avenue, Chicago, IL 60637, USA\\
$^{5}$Kavli Institute for Cosmological Physics, University of Chicago, Chicago, IL 60637, USA\\
}

\begin{document}
\label{firstpage}
\pagerange{\pageref{firstpage}--\pageref{lastpage}}
\maketitle

\begin{abstract} 
We report on the discovery of the first ultra metal-poor (UMP) star 2MASS~J20500194$-$6613298 (J2050$-$6613; \mbox{[Fe/H] = $-4.05$}) selected from the Gaia BP/RP spectral catalog that belongs to the ancient Atari disk component. We obtained a high-resolution spectrum for the star with the MIKE spectrograph on the Magellan-Clay telescope. J2050$-$6613 displays a typical chemical abundance pattern for UMP stars, including carbon and zinc enhancements. In contrast, J2050$-$6613 shows extremely high [Sr/Fe] and [Sr/Ba] ratios compared to other stars in the [Fe/H] $<-4.0$ regime. J2050$-$6613 is most likely an early Population\,II star that formed from a gas cloud that was chemically enriched by a massive Population\,III hypernova (E $> 10^{52}$\,erg). Such a Population\,III core-collapse hypernova could simultaneously explain the origin of the abundance pattern of light and heavy elements of 2MASS~J2050$-$6613 if a large amount of Sr of $\sim10^{-5}$\,M$_{\odot}$ was produced, possibly by neutrino-driven \textbf{(wind)} ejecta. Therefore, the abundance pattern of 2MASS~J2050$-$6613 places important constraints on Sr-producing nucleosynthesis sources operating in the Atari progenitor at the earliest times. 
\end{abstract}


\section{Introduction}
The very first (so-called Population\,III; hereafter Pop\,III) stars were massive and lived short \citep[e.g.,][]{pop3_massive}. No direct observation of Population\,III stars may be possible, even with the JWST telescope \citep{JWST_telescope} due to their high redshift $\geq 20$ and extremely limited apparent brightness. However, an alternative path to investigating Pop III stars is studying chemically pristine, ultra-metal poor ([Fe/H]\footnote{\metal = $\log_{10}(N_{\text{Fe}}/N_{\text{H}})_{\star}-\log_{10}(N_{\text{Fe}}/N_{\text{H}})_{\sun}$} $<-4.0$; UMP) Galactic stars. These stars preserve clean signatures of one or few enrichment events that polluted their birth clouds as discussed in a number of reviews \citep[e.g.,][]{beers&christlieb05,Frebel_Norris_2015,Frebel2018}. As such, they provide information on early star formation processes and help reconstruct the characteristics of the first stars and their nucleosynthetic yields.

Extensive efforts have been made to discover these early, most metal-poor (Pop\,II) stars over the last several decades \citep[e.g.,][]{old_EMP,frebel_EMP,SMSS_EMP,pristine_EMP}. As a result, 41 UMP stars \citep[e.g.,][]{Christlieb2002,HE1327_Nature,Caffau2011,Keller2014,Placco2021} have been confirmed with high-resolution spectroscopic  observations to date \citep{jinabase}. These searches have clearly shown UMP stars to be incredible rare, at a $\sim$1 in 1,000,000 rate \citep{Frebel2018}. This also explains why only a few dozen of stars have been found. More efficient search techniques appear to hold the key to identifying these rare objects.

Techniques have been recently refined to efficiently identify the most metal-poor candidates from narrow-band photometry \citep[e.g.,][]{pristine_MRS,SMSS_EMP,Chiti_SMSS_2020,Whitten2021,Placco2022}. This process can be further optimized by deriving [Fe/H] estimates from the new spectrophotometry data products \citep[often shortened as GaiaXP; see][]{GaiaXP1,GaiaXP2,GaiaXP3} of the third data release (DR3) of the ESA/Gaia astrometric mission \citep{Gaia_the_mission,Gaia_DR3}. Extensive efforts have been undertaken to obtain accurate [Fe/H] estimates for the $\sim220$\,million sources in the Gaia XP catalog \citep[e.g.,][]{Rix2022,Andrae2023,Yao2023,Zhang2023}. Yet, no Gaia XP UMP stars have thus far been discovered.

In a parallel effort, we have computed metallicity-sensitive synthetic photometry from the Gaia XP spectrophotometric data using {\sc{GaiaXPy}}\footnote{Available at \url{https://gaia-dpci.github.io/GaiaXPy-website}}, and utilized the grids of synthetic photometry presented in \citet{Chiti_SMSS_cat2021} to estimate [Fe/H] for sources in the Gaia XP data set (M. Mardini et al. 2024, in prep.). From high-resolution spectroscopic follow-up of our lowest [Fe/H] candidates, we report the discovery and detailed chemical abundance analysis of the relatively bright (V = 13.1\,mag) red giant star 2MASS~J20500194$-$6613298 (hereafter J2050$-$6613) with [Fe/H] = $-4.05$ and [C/Fe] = 1.24, the first confirmed UMP star from the Gaia XP catalog. We note that the low metallicity nature of J2050$-$6613 ([Fe/H] = $-4.25$) had independently been reported by \citet{SMSS_EMP}, but no high-resolution follow-up analysis was reported for the star.

\section{Target Selection, Observations, and Radial Velocity}\label{sec:obs}

The Gaia Data Processing and Analysis Consortium developed the {\sc{GaiaXPy}} package, which allows one to calculate the expected flux of an object with XP spectra through pre-loaded transmission curves. Several photometric filters were provided that are known to be sensitive to metallicities of stars, namely the SkyMapper $u,v,g,i$ filterset and the narrow-band Pristine Ca\,II\,K filter. The 360-410\,nm SkyMapper $v$-filter and the even narrower 390-400\,nm Pristine filter cover CaII\,K line at 393\,nm. Photometry through these filters can demonstrably be used to identify low-metallicity stars \citep[e.g.,][]{pristine_EMP,SMSS_EMP,Chiti_SMSS_cat2021}.

We leverage the utility of the aforementioned metallicity-sensitive filters to derive photometric metallicities for sources with Gaia XP spectra. We selected a sample of $\sim$ 61\, million Gaia XP sources that satisfy the following quality cuts: i) Galactic latitude $|b| > 10.0${\textdegree}, ii) RUWE $<$ 1.1, iii) c$\_$star $< 1.0$, and iv) E(B $-$ V) $< 0.35$. For this sample, we then derived photometric metallicities using grids of synthetic photometry that were matched to observed photometry, exactly as reported in \citet{Chiti_SMSS_cat2021}. However, we also regenerated the same grid to include the Pristine CaII\,K filter and repeated the same procedure to derive photometric metallicities with broad-band SkyMapper $g,i$ photometry. This allowed two estimates for the metallicity of each stars-- one from the SkyMapper filterset ([Fe/H]$_{SMSS}$), and the other from the Pristine + SkyMapper $g,i$ photometry ([Fe/H]$_{Pris}$). Our [Fe/H]$_{Pris}$ measurements agree to within 0.39\,dex with results from high-resolution spectroscopic [Fe/H] values, down to [Fe/H] $\sim-4.0$. For comparison, this is as nearly as good as typical medium-resolution spectroscopic results with $\sim$ 0.30\,dex uncertainties. A more detailed description of this procedure together with a catalog of metallicities will be presented in a forthcoming paper (M. Mardini et al. 2024, in prep.).

\begin{table*}
\centering
\caption{Magellan/MIKE Chemical Abundances of J2050$-$6613}
\label{tab:abund}
\begin{tabular}{lcrrrrrccccc}
\hline
Species& Method& $N$ & $\log\epsilon (\mbox{X})$ & st.dev  &st.err&  [X/H]& [X/Fe] & $\Delta T_{\text{eff}}$ & $\Delta\log(g)$ & $\Delta v_{\text{micr}}$ & $\sigma_{tot}$\\
  &     &     &       (dex)           &    & & (dex) & (dex) & $+100$ K & $+0.3$ dex & $+0.3$ \kmsec \\
\hline
C        &   Syn  &  1   &  $+$5.62 &  0.10   & 0.10 &  $-$2.81   &  $+$1.24   &  $-$0.16 &  $-$0.00 & $-$0.00 & 0.17 \\
O        &   Syn  &  1   &  $<+$6.19&  \ldots &\ldots&  $<-$2.50  &  $<+$1.55  &   \ldots &   \ldots &  \ldots &\ldots\\
Na\,I    &   EW   &  2   &  $+$1.93 &  0.08   & 0.07 &  $-$4.31   &  $-$0.26   &  $-$0.05 &  $-$0.02 & $-$0.09 & 0.12 \\
Mg\,I    &   EW   &  6   &  $+$4.12 &  0.09   & 0.04 &  $-$3.48   &  $+$0.57   &  $-$0.05 &  $-$0.06 & $-$0.12 & 0.13 \\
Al\,I    &   Syn  &  1   &  $+$1.73 &  0.10   & 0.10 &  $-$4.72   &  $-$0.67   &  $-$0.09 &  $-$0.00 & $-$0.13 & 0.17 \\
Si\,I    &   Syn  &  1   &  $+$3.60 &  0.10   & 0.10 &  $-$3.91   &  $+$0.14   &  $-$0.04 &  $-$0.01 & $-$0.03 & 0.10 \\
Ca\,I    &   EW   &  3   &  $+$2.56 &  0.09   & 0.06 &  $-$3.78   &  $+$0.27   &  $-$0.10 &  $-$0.03 & $-$0.02 & 0.11 \\
Sc\,II   &   EW   &  3   &  $-$1.34 &  0.09   & 0.05 &  $-$4.49   &  $-$0.44   &  $-$0.07 &  $-$0.08 & $-$0.03 & 0.12 \\
Ti\,I    &   EW   &  3   &  $+$1.35 &  0.13   & 0.05 &  $-$3.60   &  $+$0.45   &  $-$0.05 &  $-$0.02 & $-$0.05 & 0.09 \\
Ti\,II   &   EW   & 10   &  $+$1.11 &  0.16   & 0.05 &  $-$3.84   &  $+$0.21   &  $-$0.12 &  $-$0.07 & $-$0.11 & 0.14 \\
Cr\,I    &   EW   &  3   &  $+$1.27 &  0.07   & 0.04 &  $-$4.37   &  $-$0.32   &  $-$0.04 &  $-$0.04 & $-$0.06 & 0.09 \\
Mn\,I    &   EW   &  2   &  $+$0.20 &  0.05   & 0.05 &  $-$5.23   &  $-$1.18   &  $-$0.03 &  $-$0.02 & $-$0.07 & 0.09 \\
Fe\,I    &   EW   & 56   &  $+$3.45 &  0.16   & 0.02 &  $-$4.05   &  $+$0.00   &  $-$0.11 &  $-$0.02 & $-$0.09 & 0.13 \\
Fe\,II   &   EW   &  3   &  $+$3.46 &  0.05   & 0.04 &  $-$4.04   &  $+$0.01   &  $-$0.02 &  $+$0.08 & $-$0.05 & 0.10 \\
Co\,I    &   EW   &  8   &  $+$1.33 &  0.15   & 0.05 &  $-$3.66   &  $+$0.39   &  $-$0.02 &  $-$0.03 & $-$0.04 & 0.07 \\
Ni\,I    &   EW   & 12   &  $+$2.32 &  0.17   & 0.06 &  $-$3.90   &  $+$0.15   &  $-$0.01 &  $-$0.06 & $-$0.03 & 0.08 \\
Zn\,I    &   Syn  &  1   &  $+$1.35 &  0.10   & 0.10 &  $-$3.21   &  $+$0.95   &  $-$0.09 &  $-$0.09 & $-$0.08 & 0.12 \\
Sr\,II   &   Syn  &  2   &  $-$0.18 &  0.05   & 0.03 &  $-$3.05   &  $+$1.00   &  $-$0.10 &  $-$0.04 & $-$0.07 & 0.14 \\
Ba\,II   &   Syn  &  1   &  $-$2.95 &  0.10   & 0.10 &  $-$5.13   &  $-$1.08   &  $-$0.11 &  $-$0.08 & $-$0.11 & 0.22 \\
\hline
\end{tabular}
\\
\end{table*}

\begin{figure*}
\begin{center}
\includegraphics[clip=true,width=\textwidth]{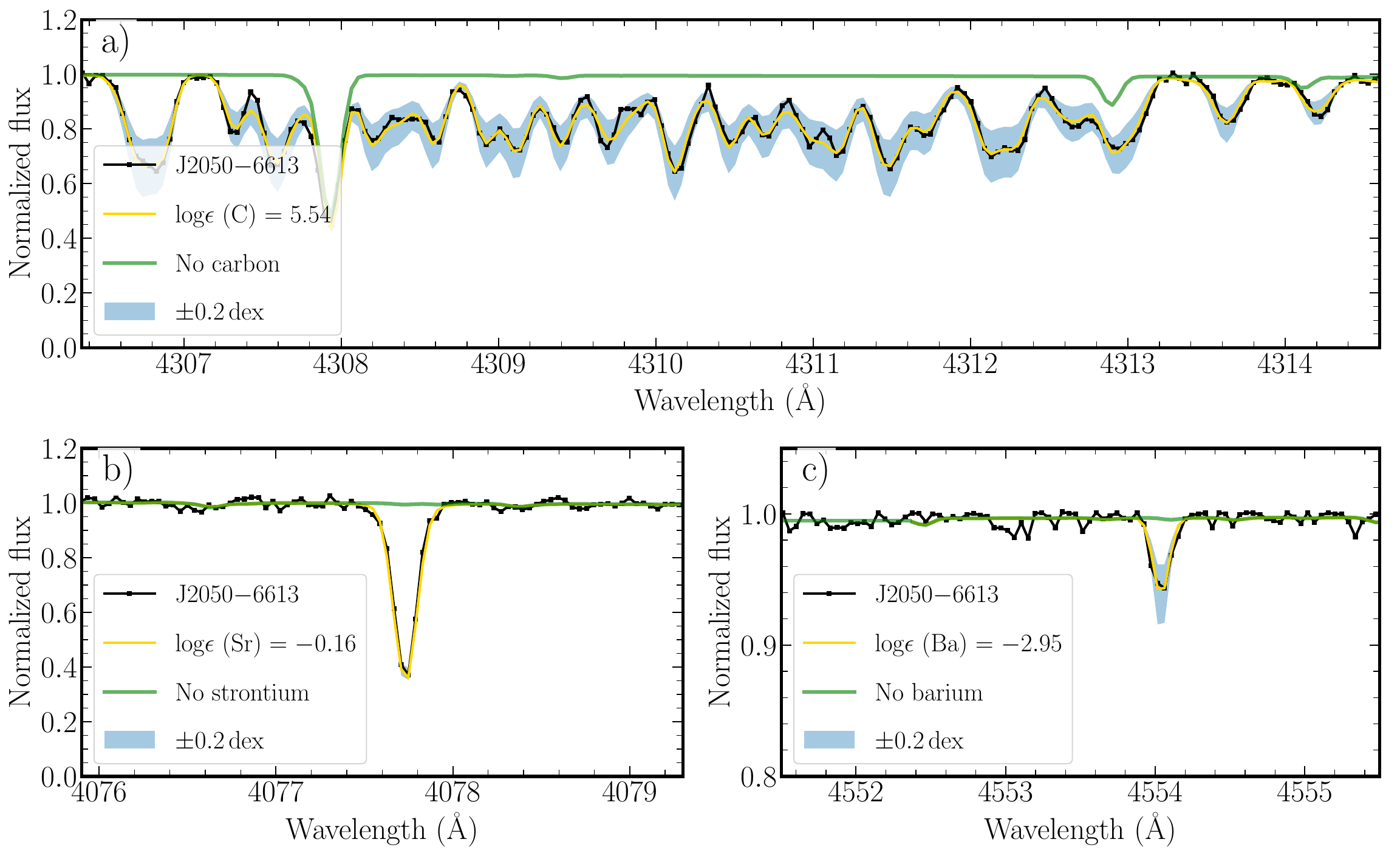} 
\caption{Portions of the observed spectrum of J2050$-$6613 used to derived chemical abundances for carbon (panel a), strontium (panel b), and barium (panel c). Black line-connected points denote the observed data. Gold lines represent the best-fit spectrum syntheses. Green lines represent syntheses with diminishing contribution by the relevant element. The shaded areas represent illustrative $\pm 0.2$\,dex variations to show corresponding abundance uncertainties. Final line abundances are listed in the legends. 
\label{fig:fig1}} 
\end{center}
\end{figure*}

To identify additional metal-poor stars with [Fe/H] $<-3.0$, we selected $\sim 25$ candidates with [Fe/H]$_{SMSS}$ $\leq -3.0$ and [Fe/H]$_{Pris}$ $\leq -3.0$, and no high-resolution spectroscopic abundances in the literature. We then obtained high-resolution spectra of 20 stars with MIKE/Magellan \citep{mike}. Given the reliability of our derived photometric metallicities, we were able to skip the time-consuming process of vetting candidates first, typically done with medium-resolution spectra and Ca\,II K line measurements. We observed J2050$-$6613 on 2023 April 22 for a total of 2000\,s using the 0\farcs7 slit. Our observation setup yielded a high spectral resolution of R $\sim$ 35,000 and 30,000 in the blue ($\lambda < 5000$) and red ($\lambda > 5000$) wavelength regime of our spectrum that covers a wide (3300\,{\AA} to 9400\,{\AA}) wavelength range. We carried out a standard data reduction procedure using the reduction pipeline developed for Magellan/MIKE observations \citep{kelson03}\footnote{Available at \url{http://obs.carnegiescience.edu/Code/python}}. The reduced data has high S/N of 110 at $\lambda \sim4000$\,{\AA}. 

We cross-correlated our MIKE spectrum against a restframe spectrum of HD122563 (a red giant with [Fe/H] $\sim-2.8$) around the Ca\,triplet region to calculate a heliocentric radial velocity. We find RV = 195.6 $\pm 0.5$\,\kms. We also retrieved RV measurements from Gaia DR3 of \citep[RV = 197.32 $\pm 2.04$\,\kmsec][]{Gaia_DR3}, from RAVE DR5 of \citep[RV = 196.50 $\pm 4.18$\,\kmsec][]{RAVE_5th}, and from RAVE DR6 of \citep[RV = 196.10 $\pm 4.28$\,\kmsec][]{RAVE_6th}. These available RV measurements for J2050$-$6613 suggests no binarity.

\section{Equivalent widths and Stellar Parameters}\label{sec:params}

We adopted an atomic and molecular data linelist generated by the linemake\footnote{Available at \url{https://github.com/vmplacco/linemake}} code \citep{Placco2021_linemake}. We measured equivalent widths (EW) for detectable lines (see Table~\ref{tab:abund}) in our high-resolution spectrum using the Spectroscopy Made Hard ({\sc{SMH}}) software \citep{casey14}\footnote{Available at \url{https://github.com/andycasey/smhr}} by fitting Gaussian profiles to the absorption features. We note that the selected lines for the EW measurements were checked for no carbon contamination. Also, we visually examined each of the lines for potential blending and corrected the fits for any unreliable continuum placements. 

We determined the atmospheric stellar parameters for J2050$-$6613 using a combination of photometric and spectroscopic approaches, similarly to what was presented in \citet{Mardini23_dwarf}. We estimate the photometric effective temperature ($T_{\rm{eff}}$) using accurate G, BP and RP magnitudes for our star, as reported in the source catalog published by \citet{Gaia_DR3}. We estimated the bolometric correction following \citet{Casagrande2018BC1,Casagrande2018BC2}. We used the best polynomial fit of the color–$T_{\rm{eff}}$ from \citet{Mucciarelli2021} to derive $T_{\rm{eff}}$. To obtain the median $T_{\rm{eff}}$ and its uncertainty, we generated 10,000 random realizations drawn from each input parameter and their corresponding uncertainties. This way, we calculated our final $T_{\rm{eff}}$ = 5000 $\pm$ 100\,K which is the median value of the distribution and its standard deviation. 

We used our EW measurements, the latest version of MOOG \citep{moog} \footnote{Available at \url{https://github.com/alexji/moog17scat}}, and one-dimensional plane-parallel model atmospheres with $\alpha$-enhancement \citep{Castelli2004} wrapped within {\sc{SMH}} to derive LTE chemical abundances of Fe\,I and Fe\,II lines. We then used the Fe\,I and Fe\,II abundances to iteratively constrain the surface gravity ($\log g$) and microturbulence (v$_{\text{micro}}$) by adopting $\log g$ that show agreement between the Fe\,I and Fe\,II mean abundances and enforcing no trends between the reduced EW and Fe\,I abundances. Our final stellar parameters are $T_{\rm{eff}}$ = 5000 $\pm$ 100\,K, $\log g$= 1.80 $\pm$ 0.30\,dex, v$_{\text{micro}}$= 1.40 $\pm$ 0.30\,\kms, and [Fe/H] = $-$4.05 $\pm$ 0.16\,dex. We note that we adopted typical uncertainties for $\log g$ and v$_{\text{micro}}$. Also, we adopted the standard deviation of the Fe\,I abundances as the uncertainty for [Fe/H]. For completeness, we also obtained stellar parameters from a traditional spectroscopic approach and found $T_{\rm{eff}}$ = 4350\,K, $\log g$ = 0.21\,dex, v$_{\text{micro}}$ = 1.86\,\kmsec, and [Fe/H] = $-4.75$\,dex. It is well established that spectroscopic temperatures are cooler by several hundred degrees than photometrically derived ones. Correcting the spectroscopic temperature following the approach presented in \citet{frebel13} would yield $T_{\rm{eff}}$ = 4585\,K, $\log g$ = 0.79\,dex, v$_{\text{micro}}$ = 1.52\,\kmsec, and [Fe/H] = $-4.48$\,dex. However, no straightforward  explanation for the large ($\sim415$\,K) difference between our derived temperatures can be drawn, but note that these temperatures yield [Fe/H] $\lesssim-$4.0. Thus, J2050$-$6613 is truly a UMP star.

\section{Chemical abundances}\label{sec:chem}

We determined elemental abundance ratios ([X/Fe]) from both spectrum synthesis and EW analysis of absorption lines detected in our Magellan/MIKE spectrum. We adopted solar abundances ($\log\epsilon (\mbox{X})_{\odot}$) from \citet{asplund09}. A summary of the EW fitting procedure and analysis is provided in Section~\ref{sec:params}. For the spectrum synthesis approach, we generated synthetic spectra to match the observational data to derive chemical abundances for seven elements (see the second column of Table~\ref{tab:abund}). Figure~\ref{fig:fig1} shows an illustrative example of our spectral synthesis fitting. We took into consideration the effects of hyperfine splitting when determining the abundances from the Sr\,II and Ba\,II lines, as discussed in \citet{Bergemann2012} and \citet{Gallagher2010}, respectively. Our final abundances are listed in Table~\ref{tab:abund}. We also used the strongest line of the O\,I triplet $\sim$ 7770\,{\AA} to place an upper limit on the oxygen abundance. 
We determined reasonable line uncertainties for the elements derived from spectral synthesis (C, Al, Si, Zn, Sr, and Ba) by comparing a range of synthetic spectra that were consistent with the continuum (see e.g., the shaded area in Figure~\ref{fig:fig1}).

The fifth column presents the standard deviation of the mean abundances as derived from individual line measurements. Note that for elements with few available lines ($N$<4; e.g., Na), the standard deviations were adjusted to account for the small number samples, following \citet{Keeping62}. This prevents uncertainties from being underestimated due to small number statistics. To arrive at these adjusted values, we use $\sigma$ = Rk, where k is a factor computed for each N derived to approximate Gaussian statistics for small N samples. Values of k can be found in table~2.5 of \citet{Keeping62}. R is the maximum range covered by individual measurements. In the sixth column, we report the standard error, based on st.err = Rk$^{\star}$, and again, corrected for small number samples. The k$^{\star}$ values are taken from table~2.5 of \citet{Keeping62}.

We also estimated the effects of the uncertainty in our stellar parameters on the derived atmospheric abundances. The ninth ($\Delta T_{\text{eff}}$), tenth ($\Delta\log(g)$), and eleventh ($\Delta v_{\text{micr}}$) columns list the systematic uncertainties due to changes in one stellar parameter within their uncertainties ($\Delta T_{\rm{eff}}$ = 100\,K, $\Delta \log g$ = 0.3\,dex, and $\Delta$ v$_{\text{micro}}$ = 0.3\,\kmsec). We finally calculated the total uncertainties as quadratic sums of each uncertainty estimate, using our calculated standard errors. 

Overall, J2050$-$6613 is a carbon enhanced ([C/Fe] = 1.24) UMP ([Fe/H] = $-4.05$) star, adding to the overwhelming number of carbon-rich halo stars at the lowest metallicities. An evolutionary correction of [C/Fe] = $+0.08$, based on \citet{placco14_carbon}, is applied here. The other abundances of J2050$-$6613 also agree well with those of other known UMP stars (i.e., enhancements in $\alpha$-elements and Zn relative to iron). The neutron-capture element abundances are somewhat unusual in J2050$-$6613. The Ba abundance is low ([Ba/Fe] $= -1.04$), as it is typical for other halo stars at [Fe/H] $\sim-4$. However, the Sr abundance, of [Sr/Fe] $= +0.95$, is among the highest of UMP stars. This leads to one of the largest known [Sr/Ba] ratios ([Sr/Ba] = 1.99). Note that using the corrected spectroscopic stellar parameters would yield [Sr/Fe] $= +1.10$ and [Ba/Fe] $= -1.30$. We illustrate this behavior in panels a and b of Figure~\ref{fig:fig2} where we show [Sr/H] and [Ba/H] as a function of [Fe/H]. Panel c of Figure~\ref{fig:fig2} shows [Sr/Ba] versus [Ba/Fe]. The blue asterisk denotes the location of our star, and orange points are chemical abundances data for metal-poor stars compiled in \citet{jinabase}. Some relevant comparison metal-poor stars (as discussed in Section~\ref{sec:discuss}) are plotted with symbols as listed in the legend.

\section{Possible Origin Scenarios for J2050$-$6613}\label{sec:discuss}

We now utilize the observed chemical abundance signature of J2050$-$6613 to gain insights into its birth environment and how its progenitor gas cloud was chemically enriched by the first stars. The empirical transition discriminant ($D_{\text{trans}}$) criterion from \citet{dtrans} employs carbon and oxygen abundances to set a limit for sufficient fragmentation for early low-mass Pop\,II star formation through C and O fine-structure line cooling. 

To calculate $D_{\text{trans}}$ for J2050$-$6613, we used our measurement of [C/H] and the upper limit on [O/H]. Since we only have an upper limit for O we cannot derive a $D_{\text{trans}}$ value that would meaningfully suggest whether it is above the threshold of $D_{\text{trans}}$ $ = -3.50$.
However, given that our [C/O] $> -$0.31 agrees with the suggested range of $-0.7 <$ [C/O] $ < 0.2$ by \citet{fn13} for metal-poor stars, we adopt [C/O] = 0.2 to estimate $D_{\text{trans}}$. This way we obtain $D_{\text{trans}}\gtrsim -2.8$.
This reasonable lower limit is already well above the threshold. 
We note that if we were to apply any NLTE corrections to the O triplet, the O abundance would decrease even further and effectively not change the $D_{\text{trans}}$ value. 
This suggests that the composition of the birth gas cloud fragmented sufficiently for low-mass stars to form, and in the case of J2050$-$6613 to form with a high natal C abundance.

For CEMP-no stars ([C/Fe] $> 0.7$ and [Ba/Fe] $< 0.0$), such as J2050$-$6613, \citet{Rossi_CEMPno_pop3} suggested a criterion for identifying their origin based on their carbon, magnesium and iron abundances. A similar approach was suggested by \citet{Tilman_mono} although they did not specifically consider CEMP-no stars. J2050$-$6613 has A(C) = 5.62 and [C/Mg] = 0.67, which indicates predominant prior enrichment by massive metal-free stars to the local birth environment \citep[e.g.,][]{placco16,Almusleh2021,Mardini_2019b}, likely by those that produced a lot of carbon.

\begin{figure*}
\begin{center}
\includegraphics[clip=true,width=\textwidth]{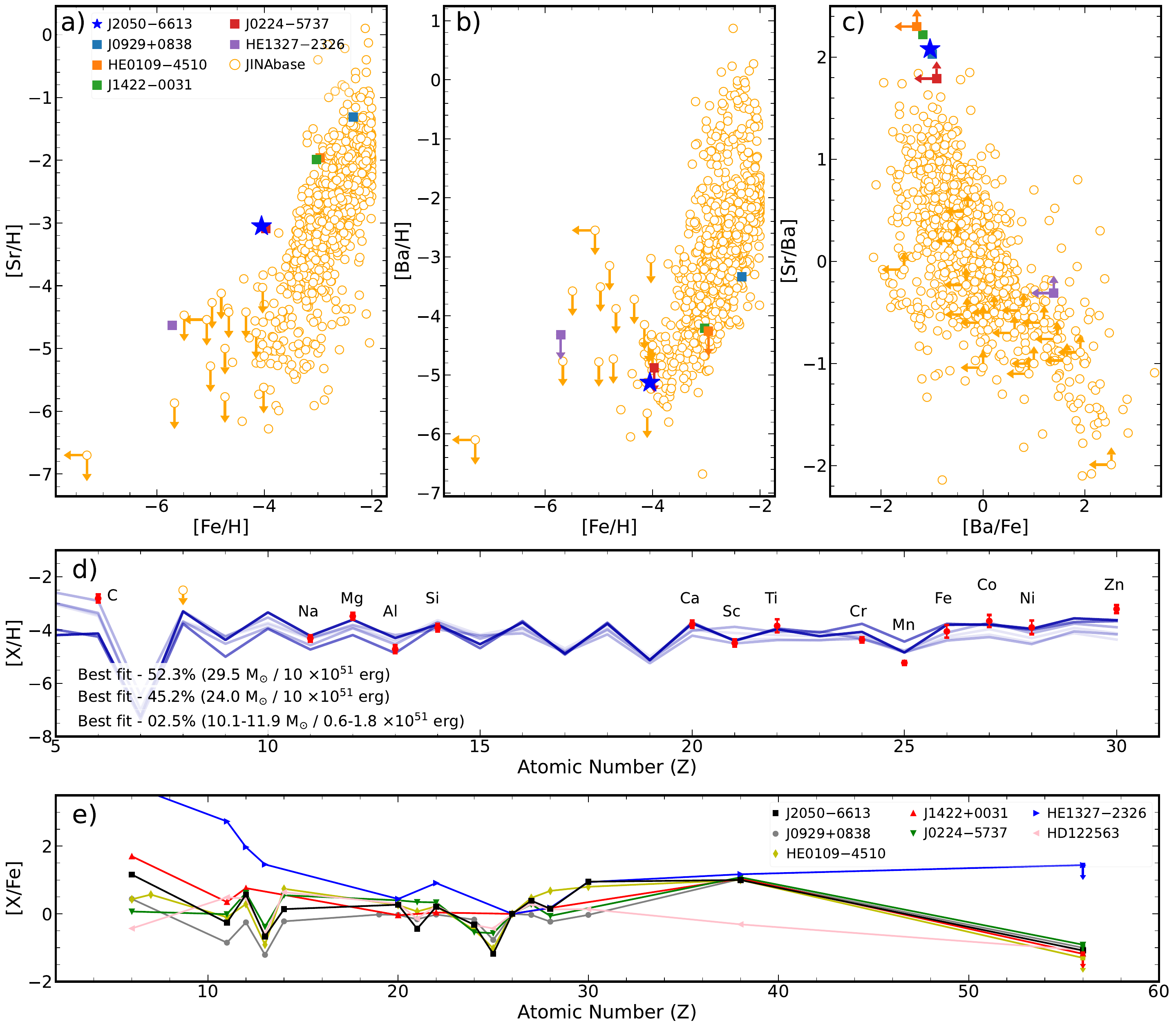} 
\caption{Observed abundance trends of [Sr/H] (panel a) and [Ba/H] (panel b) as a function of [Fe/H]. Panel c represents the [Sr/Ba] versus [Ba/Fe]. The blue asterisk denotes the  abundances for J2050$-$6613. Gray points represent the data for metal-poor stars collected by JINAbase. Other noteworthy metal-poor stars are indicated with various symbols. Solid lines in panel d show the best-fit Pop\,III theoretical yields (with progenitor properties) from \citet{heger_woosley10} and color-coded by their fractional occurrence. Panel e shows LTE elemental abundance patterns for J2050$-$6613 and the other noteworthy metal-poor stars (see the legend). Metal-poor star data taken from: 
\citet{roederer14c}; \citet{barklem05}; \citet{Ryan1991}; \citet{Cohen2004}; \citet{Hollek2011}; \citet{Bonifacio2012}; \citet{Mardini_2019a,Mardini_2019b,Mardini_2020,Mardini2022b}; \citet{Aguado2017b}; \citet{Yong2013}; \citet{Frebel2010}; \citet{Hansen2015}; \citet{Jacobson2015}; \citet{ezzeddine19}; \citet{Spite2014}; \citet{cayrel2004}; \citet{Norris2007}; \citet{francois07}; \citet{Aguado2021}; \citet{Lai2004}; \citep{Mashonkina2017}; \citet{Almusleh2021}; \citet{Placco2020}; \citet{Ryan1996}; \citet{Aguado2018}; \citet{Aoki2007}; \citet{Aoki2013}; \citet{Cohen2013}; \citet{Honda2011}; \citet{Lai2008}; \citet{Casey2015}; \citet{Masseron2006}; \citet{Rich2009}; \citet{Ryan1999}; \citet{Spite1999}; \citet{Depagne2000}; \citet{Spite2000}; \citet{Sivarani2006}; \citet{Norris2001}; \citet{Mardini2019c}; \citet{Hansen2018_Sgr}; \citet{placco2014a}; \citet{Caffau2011a}; \citet{HE1327_Nature}; \citet{frebel08}; \citet{Behara2010}; \citet{Caffau2011}; \citet{Aguado2017}; \citet{Carretta2002}; \citet{For2010}; \citet{Keller2014}; \citet{frebel15b}; \citet{Caffau2013}; \citet{Plez2005}; \citet{Christlieb2002}; \citet{Frebel2019}; \citet{placco2023}. \label{fig:fig2}}
\end{center}
\end{figure*}

Assuming J2050$-$6613 is a second-generation star, then its atmospheric abundance pattern can be expected to reflect the chemical yields of the Pop\,III progenitor. To constrain the progenitor characteristics, we matched 10,000 abundance patterns, generated from normal Gaussian distributions of our derived abundances and uncertainties, to theoretical Pop\,III models taken from \citet{heger_woosley10} and using the {\sc{StarFit}}\footnote{Available at \url{http://starfit.org}} code. We find a reasonable fit for most elements, although the observed carbon abundance often appears to be underestimated by the models. The best-fit parameters of $\sim 97\%$ of our generated patterns suggest a massive progenitor (29.5 and 24.0\,M$\odot$, respectively, as seen in panel d of Figure~\ref{fig:fig2}). These models have high explosion energies of E = 10$^{52}$\,erg. Such energetic hypernovae most likely enriched the birth gas cloud of J2050$-$6613. In contrast, few models (2.5\%) have lower masses (M = $10$ - $11$\,M$_{\odot}$). They match the observed [C/Fe] for our star but consistently underestimate [Zn/Fe]. Zinc is produced in the complete Si-burning shell deep in the star. 
Consequently, the production of a substantial amount of Zn relies on the energy released during an explosion and is commonly associated with hypernovae \citep{umeda&nomoto05}. Along with Zn enhancements, Pop\,III hypernova models also predict an enhancement in Co alongside lower values for Mn and Cr \citep{umeda&nomoto05}. 

Qualitatively, J2050$-$6613 follows this principal pattern. However, the hypernova models still somewhat underproduce Zn by about 0.3\,dex as well as carbon. In contrast, the lower mass progenitor models cannot produce nearly enough Zn which we take as a benchmark for constraining the progenitor. We thus adopt the properties of a high-mass hypernovae for the Pop\,III progenitor. We note that it is common for some of the individual abundances to be fit poorly; as discussed by \citet{heger_woosley10} and also \citet{Magg2020_fits} this is at least partially due to uncertainties in our observations (i.e., stellar parameters and atmospheric abundances), theoretical models (e.g., yield predictions), as well as the fitting algorithm. Keeping these caveats in mind, and notwithstanding, this comparison provides a helpful indication of the potential stellar mass and explosion energy of the progenitor.

We now discuss possible origins of the observed neutron-capture elements. J2050$-$6613 exhibits a strong enhancement in Sr ([Sr/Fe] = 0.95), in stark contrast to its much more ``typical'' Ba ([Ba/Fe] = $-$1.04) abundance. Among the other known UMP stars, only HE~1327$-$2326 ([Fe/H] $= -5.71$; [Sr/Fe] = 1.08; [Ba/Fe] $<$ 1.39; \citealt{frebel08}) and J0224$-$5737 ([Fe/H] $= -3.97$; [Sr/Fe] = 1.08; [Ba/Fe] $<$ -0.91; \citealt{Jacobson2015}) have similar Sr enhancement levels; however, these stars only have Ba upper limits available. For comparison, the Sr abundances of other UMP halo stars (e.g., \citealt{frebel14}, \citealt{Andales_SASS}) typically have much lower values as low as [Sr/Fe] $\sim-1.8$ at [Fe/H] $\sim-4.0$ which is nearly three dex lower than what we observed in J2050$-$6613.

However, there are several other stars with large [Sr/Ba] ratios which we highlight in figure~\ref{fig:fig2}: J0929+0838 ([Fe/H] = $-2.34$; \citealt{Ezzeddine_RPA_2020}), HE0109$-$4510 ([Fe/H] = $-2.96$; \citealt{Hansen2015}), J1422+0031 ([Fe/H] = $-3.03$; \citealt{Aoki2013}), and, again, J0224$-$5737 ([Fe/H] = $-3.97$; \citealt{Jacobson2015}). Note that the latter two stars only have upper limits on their Ba abundances, implying lower limits on their [Sr/Ba] values. Nevertheless, all these objects cluster above all other stars that form the main observed trend, thus opening up the parameter space. For completeness, we also add HE~1327$-$2326; note that its high upper limit of [Ba/Fe] $<$ 1.39 renders its low [Sr/Ba] $ > -0.31$ lower limit largely meaningless.

Sr abundances in combination with significantly lower Ba values are often attributed to the operation of the limited-$r$ process \citep{Frebel2018}. However, in comparison to the representative limited-$r$ star HD122563 \citep{Honda2011}, J2050$-$6613 shows a much higher relative Sr abundance as well as a more extreme [Sr/Ba] ratio (HD122563 has [Sr/Ba] $\sim$0.8\,dex). We added HD122563 to panel e of Figure~\ref{fig:fig2} for comparison. Altogether, it appears that there is a new class of objects with such high [Sr/Ba] ratios of $\gtrsim$ 2. We plot the stars common element abundances in panel e of Figure~\ref{fig:fig2}. As can be seen, all stars have qualitatively very similar patterns. Possible exception may be some variations in the Ba abundances given that several stars only have upper limits at present.  

To learn more about these stars and their nucleosynthetic origins, we now discuss core-collapse supernovae (CCSN) as a major source of neutron-capture elements \citep[e.g.,][]{Sr_from_CCSN2,Arcones2007,Wanajo2011}. Specifically, we attempt to derive yield predictions for Sr and Ba. Assuming that J2050$-$6613 formed from a gas cloud with $10{^5}$\,M$_{\odot}$, the observed abundances imply a large Sr yield of $\sim10^{-5}$\,M$_{\odot}$ alongside a Ba yield of only $\sim10^{-8}$\,M$_{\odot}$. Whether these very different yields are physically plausible to be produced within one massive first star remains to be seen. In any case, it appears that CCSN must produce extremely variable Sr yields as reflected in the different [Fe/H] levels observed. Adding rotation to the models can, however, enhance the Sr production \citep[e.g.,][]{Frischknecht_Sr_Ba}. Also, neutrino-driven winds that follows the successful CCSN might synthesize additional Sr \citep[e.g.,][]{Bliss_neutrino-driven}. Interestingly, our suggested Sr yield appears to (coincidentally) be similar to what the kilonova AT2017gfo produced \citep{Watson_Sr_detection}. However, the coalescence time of binary neutron star mergers argues against a compact object merger as the origin of Sr observed in our star. However, if CCSN yields are found to produce only a narrow range of Sr yields, the new class of high [Sr/Ba] stars suggests a likely contribution from an additional source (with the possible caveat that this source should produce negligible amounts of iron as to not overproduce the stellar [Fe/H] of the initial/main source). A discussion of Sr yields and associated astrophysical sites, can be found in \citet{Hansen2013}.

In summary, the abundance signature of J2050$-$6613 suggests that its Pop\,III progenitor was massive and very energetic, and that it must have produced copious amounts of Sr but little Ba, possibly through a limited r-process while also having a low Fe yield to account for the UMP nature of the star. However, other scenarios, possibly involving two sources or sites, should be further explored to place new constraints on the nucleosynthetic history of this class of very low metallicity, high [Sr/Ba] stars, including J2050$-$6613. 
Interestingly, since J2050$-$6613 is also another confirmed ultra-metal-poor member of the so-called Atari disk \citep{Mardini2022}, this suggests that significant C and Sr production and/or the limited $r$-process may have been a major pathways for element nucleosynthesis in the progenitor system of the Atari component. Atari itself then likely formed from a very early ($\geq 9$\,Gyr) radial merger event of that progenitor that entered the proto-galactic disk to form a component separate from the canonical thick disk. For more details, interested readers are referred to \citet{Mardini2022}. In any case, the Atari disk appears to contain a significant amount of the most metal-poor stars, including five stars with [Fe/H] $<-$4.0. J2050$-$6613 adds to this growing number of truly ancient UMP stars of accreted second generation stars present in the Galactic disk.


\section*{Acknowledgements}
We thank Vinicius Placco, Ian Roederer, and Shinya Wanajo for helpful discussions. This work was initiated at an event supported by the National Science Foundation under Grant No. PHY-1430152 (JINA Center for the Evolution of the Elements). This work is supported by Basic Research Grant (Super AI) of Institute for AI and Beyond of the University of Tokyo. M.K.M. acknowledges partial support from NSF grant OISE 1927130 (International Research Network for Nuclear Astrophysics/IReNA). A.F. acknowledges support from  NSF grants AST-1716251 and AST-2307436. A.C. is supported by a Brinson Prize Fellowship at the University of Chicago/KICP.

This work has made use of data from the European Space Agency (ESA) mission
{\it Gaia} (\url{https://www.cosmos.esa.int/gaia}), processed by the {\it Gaia}
Data Processing and Analysis Consortium (DPAC,
\url{https://www.cosmos.esa.int/web/gaia/dpac/consortium}). Funding for the DPAC
has been provided by national institutions, in particular the institutions
participating in the {\it Gaia} Multilateral Agreement.

\section*{Data Availability}
The individual line measurements are provided as supplementary
material. The reduced spectra can be obtained by reasonable request
to the corresponding author.

\bibliographystyle{mnras}
\bibliography{references}

\begin{thebibliography}{}
\makeatletter
\relax
\def\mn@urlcharsother{\let\do\@makeother \do\$\do\&\do\#\do\^\do\_\do\%\do\~}
\def\mn@doi{\begingroup\mn@urlcharsother \@ifnextchar [ {\mn@doi@}
  {\mn@doi@[]}}
\def\mn@doi@[#1]#2{\def\@tempa{#1}\ifx\@tempa\@empty \href
  {http://dx.doi.org/#2} {doi:#2}\else \href {http://dx.doi.org/#2} {#1}\fi
  \endgroup}
\def\mn@eprint#1#2{\mn@eprint@#1:#2::\@nil}
\def\mn@eprint@arXiv#1{\href {http://arxiv.org/abs/#1} {{\tt arXiv:#1}}}
\def\mn@eprint@dblp#1{\href {http://dblp.uni-trier.de/rec/bibtex/#1.xml}
  {dblp:#1}}
\def\mn@eprint@#1:#2:#3:#4\@nil{\def\@tempa {#1}\def\@tempb {#2}\def\@tempc
  {#3}\ifx \@tempc \@empty \let \@tempc \@tempb \let \@tempb \@tempa \fi \ifx
  \@tempb \@empty \def\@tempb {arXiv}\fi \@ifundefined
  {mn@eprint@\@tempb}{\@tempb:\@tempc}{\expandafter \expandafter \csname
  mn@eprint@\@tempb\endcsname \expandafter{\@tempc}}}

\bibitem[\protect\citeauthoryear{{Abel}, {Bryan}  \& {Norman}}{{Abel}
  et~al.}{2002}]{pop3_massive}
{Abel} T.,  {Bryan} G.~L.,   {Norman} M.~L.,  2002, \mn@doi [Science]
  {10.1126/science.295.5552.93}, \href
  {https://ui.adsabs.harvard.edu/abs/2002Sci...295...93A} {295, 93}

\bibitem[\protect\citeauthoryear{{Abohalima} \& {Frebel}}{{Abohalima} \&
  {Frebel}}{2018}]{jinabase}
{Abohalima} A.,  {Frebel} A.,  2018, \mn@doi [\apjs]
  {10.3847/1538-4365/aadfe9}, \href
  {https://ui.adsabs.harvard.edu/abs/2018ApJS..238...36A} {238, 36}

\bibitem[\protect\citeauthoryear{{Aguado}, {Allende Prieto}, {Gonz{\'a}lez
  Hern{\'a}ndez}, {Rebolo}  \& {Caffau}}{{Aguado} et~al.}{2017a}]{Aguado2017b}
{Aguado} D.~S.,  {Allende Prieto} C.,  {Gonz{\'a}lez Hern{\'a}ndez} J.~I.,
  {Rebolo} R.,   {Caffau} E.,  2017a, \mn@doi [\aap]
  {10.1051/0004-6361/201731320}, \href
  {https://ui.adsabs.harvard.edu/abs/2017A&A...604A...9A} {604, A9}

\bibitem[\protect\citeauthoryear{{Aguado}, {Gonz{\'a}lez Hern{\'a}ndez},
  {Allende Prieto}  \& {Rebolo}}{{Aguado} et~al.}{2017b}]{Aguado2017}
{Aguado} D.~S.,  {Gonz{\'a}lez Hern{\'a}ndez} J.~I.,  {Allende Prieto} C.,
  {Rebolo} R.,  2017b, \mn@doi [\aap] {10.1051/0004-6361/201730654}, \href
  {https://ui.adsabs.harvard.edu/abs/2017A&A...605A..40A} {605, A40}

\bibitem[\protect\citeauthoryear{{Aguado}, {Allende Prieto}, {Gonz{\'a}lez
  Hern{\'a}ndez}  \& {Rebolo}}{{Aguado} et~al.}{2018}]{Aguado2018}
{Aguado} D.~S.,  {Allende Prieto} C.,  {Gonz{\'a}lez Hern{\'a}ndez} J.~I.,
  {Rebolo} R.,  2018, \mn@doi [\apjl] {10.3847/2041-8213/aaadb8}, \href
  {https://ui.adsabs.harvard.edu/abs/2018ApJ...854L..34A} {854, L34}

\bibitem[\protect\citeauthoryear{{Aguado} et~al.,}{{Aguado}
  et~al.}{2019}]{pristine_MRS}
{Aguado} D.~S.,  et~al., 2019, \mn@doi [\mnras] {10.1093/mnras/stz2643}, \href
  {https://ui.adsabs.harvard.edu/abs/2019MNRAS.490.2241A} {490, 2241}

\bibitem[\protect\citeauthoryear{{Aguado} et~al.,}{{Aguado}
  et~al.}{2021}]{Aguado2021}
{Aguado} D.~S.,  et~al., 2021, \mn@doi [\mnras] {10.1093/mnras/staa3250}, \href
  {https://ui.adsabs.harvard.edu/abs/2021MNRAS.500..889A} {500, 889}

\bibitem[\protect\citeauthoryear{{Almusleh}, {Taani}, {{\"O}zdemir}, {Rah},
  {Al-Wardat}, {Zhao}  \& {Mardini}}{{Almusleh} et~al.}{2021}]{Almusleh2021}
{Almusleh} N.~A.,  {Taani} A.,  {{\"O}zdemir} S.,  {Rah} M.,  {Al-Wardat}
  M.~A.,  {Zhao} G.,   {Mardini} M.~K.,  2021, \mn@doi [Astronomische
  Nachrichten] {10.1002/asna.202113867}, \href
  {https://ui.adsabs.harvard.edu/abs/2021AN....342..625A} {342, 625}

\bibitem[\protect\citeauthoryear{{Andales}, {Figueiredo}, {Fienberg}, {Mardini}
   \& {Frebel}}{{Andales} et~al.}{2023}]{Andales_SASS}
{Andales} H.~D.,  {Figueiredo} A.~S.,  {Fienberg} C.~G.,  {Mardini} M.~K.,
  {Frebel} A.,  2023, \mnras

\bibitem[\protect\citeauthoryear{{Andrae}, {Rix}  \& {Chandra}}{{Andrae}
  et~al.}{2023}]{Andrae2023}
{Andrae} R.,  {Rix} H.-W.,   {Chandra} V.,  2023, \mn@doi [arXiv e-prints]
  {10.48550/arXiv.2302.02611}, \href
  {https://ui.adsabs.harvard.edu/abs/2023arXiv230202611A} {p. arXiv:2302.02611}

\bibitem[\protect\citeauthoryear{{Aoki}, {Beers}, {Christlieb}, {Norris},
  {Ryan}  \& {Tsangarides}}{{Aoki} et~al.}{2007}]{Aoki2007}
{Aoki} W.,  {Beers} T.~C.,  {Christlieb} N.,  {Norris} J.~E.,  {Ryan} S.~G.,
  {Tsangarides} S.,  2007, \mn@doi [\apj] {10.1086/509817}, \href
  {https://ui.adsabs.harvard.edu/abs/2007ApJ...655..492A} {655, 492}

\bibitem[\protect\citeauthoryear{{Aoki} et~al.,}{{Aoki}
  et~al.}{2013}]{Aoki2013}
{Aoki} W.,  et~al., 2013, \mn@doi [\aj] {10.1088/0004-6256/145/1/13}, \href
  {https://ui.adsabs.harvard.edu/abs/2013AJ....145...13A} {145, 13}

\bibitem[\protect\citeauthoryear{{Arcones}, {Janka}  \& {Scheck}}{{Arcones}
  et~al.}{2007}]{Arcones2007}
{Arcones} A.,  {Janka} H.~T.,   {Scheck} L.,  2007, \mn@doi [\aap]
  {10.1051/0004-6361:20066983}, \href
  {https://ui.adsabs.harvard.edu/abs/2007A&A...467.1227A} {467, 1227}

\bibitem[\protect\citeauthoryear{{Asplund}, {Grevesse}, {Sauval}  \&
  {Scott}}{{Asplund} et~al.}{2009}]{asplund09}
{Asplund} M.,  {Grevesse} N.,  {Sauval} A.~J.,   {Scott} P.,  2009, \mn@doi
  [\araa] {10.1146/annurev.astro.46.060407.145222}, \href
  {https://ui.adsabs.harvard.edu/abs/2009ARA&A..47..481A} {47, 481}

\bibitem[\protect\citeauthoryear{{Barklem} et~al.,}{{Barklem}
  et~al.}{2005}]{barklem05}
{Barklem} P.~S.,  et~al., 2005, \mn@doi [\aap] {10.1051/0004-6361:20052967},
  \href {https://ui.adsabs.harvard.edu/abs/2005A&A...439..129B} {439, 129}

\bibitem[\protect\citeauthoryear{{Beers} \& {Christlieb}}{{Beers} \&
  {Christlieb}}{2005}]{beers&christlieb05}
{Beers} T.~C.,  {Christlieb} N.,  2005, \mn@doi [\araa]
  {10.1146/annurev.astro.42.053102.134057}, \href
  {http://adsabs.harvard.edu/abs/2005ARA%26A..43..531B} {43, 531}

\bibitem[\protect\citeauthoryear{{Behara}, {Bonifacio}, {Ludwig}, {Sbordone},
  {Gonz{\'a}lez Hern{\'a}ndez}  \& {Caffau}}{{Behara}
  et~al.}{2010}]{Behara2010}
{Behara} N.~T.,  {Bonifacio} P.,  {Ludwig} H.~G.,  {Sbordone} L.,
  {Gonz{\'a}lez Hern{\'a}ndez} J.~I.,   {Caffau} E.,  2010, \mn@doi [\aap]
  {10.1051/0004-6361/200913213}, \href
  {https://ui.adsabs.harvard.edu/abs/2010A&A...513A..72B} {513, A72}

\bibitem[\protect\citeauthoryear{{Bergemann}, {Hansen}, {Bautista}  \&
  {Ruchti}}{{Bergemann} et~al.}{2012}]{Bergemann2012}
{Bergemann} M.,  {Hansen} C.~J.,  {Bautista} M.,   {Ruchti} G.,  2012, \mn@doi
  [\aap] {10.1051/0004-6361/201219406}, \href
  {https://ui.adsabs.harvard.edu/abs/2012A&A...546A..90B} {546, A90}

\bibitem[\protect\citeauthoryear{{Bernstein}, {Shectman}, {Gunnels},
  {Mochnacki}  \& {Athey}}{{Bernstein} et~al.}{2003}]{mike}
{Bernstein} R.,  {Shectman} S.~A.,  {Gunnels} S.~M.,  {Mochnacki} S.,   {Athey}
  A.~E.,  2003, in {Iye} M.,  {Moorwood} A. F.~M.,  eds,  Society of
  Photo-Optical Instrumentation Engineers (SPIE) Conference Series Vol. 4841,
  Instrument Design and Performance for Optical/Infrared Ground-based
  Telescopes. pp 1694--1704, \mn@doi{10.1117/12.461502}

\bibitem[\protect\citeauthoryear{{Bliss}, {Arcones}, {Montes}  \&
  {Pereira}}{{Bliss} et~al.}{2020}]{Bliss_neutrino-driven}
{Bliss} J.,  {Arcones} A.,  {Montes} F.,   {Pereira} J.,  2020, \mn@doi [\prc]
  {10.1103/PhysRevC.101.055807}, \href
  {https://ui.adsabs.harvard.edu/abs/2020PhRvC.101e5807B} {101, 055807}

\bibitem[\protect\citeauthoryear{{Bonifacio}, {Sbordone}, {Caffau}, {Ludwig},
  {Spite}, {Gonz{\'a}lez Hern{\'a}ndez}  \& {Behara}}{{Bonifacio}
  et~al.}{2012}]{Bonifacio2012}
{Bonifacio} P.,  {Sbordone} L.,  {Caffau} E.,  {Ludwig} H.~G.,  {Spite} M.,
  {Gonz{\'a}lez Hern{\'a}ndez} J.~I.,   {Behara} N.~T.,  2012, \mn@doi [\aap]
  {10.1051/0004-6361/201219004}, \href
  {https://ui.adsabs.harvard.edu/abs/2012A&A...542A..87B} {542, A87}

\bibitem[\protect\citeauthoryear{{Caffau} et~al.,}{{Caffau}
  et~al.}{2011a}]{Caffau2011}
{Caffau} E.,  et~al., 2011a, \mn@doi [\nat] {10.1038/nature10377}, \href
  {https://ui.adsabs.harvard.edu/abs/2011Natur.477...67C} {477, 67}

\bibitem[\protect\citeauthoryear{{Caffau} et~al.,}{{Caffau}
  et~al.}{2011b}]{Caffau2011a}
{Caffau} E.,  et~al., 2011b, \mn@doi [\aap] {10.1051/0004-6361/201117530},
  \href {https://ui.adsabs.harvard.edu/abs/2011A&A...534A...4C} {534, A4}

\bibitem[\protect\citeauthoryear{{Caffau} et~al.,}{{Caffau}
  et~al.}{2013}]{Caffau2013}
{Caffau} E.,  et~al., 2013, \mn@doi [\aap] {10.1051/0004-6361/201322213}, \href
  {https://ui.adsabs.harvard.edu/abs/2013A&A...560A..15C} {560, A15}

\bibitem[\protect\citeauthoryear{{Carrasco} et~al.,}{{Carrasco}
  et~al.}{2021}]{GaiaXP1}
{Carrasco} J.~M.,  et~al., 2021, \mn@doi [\aap] {10.1051/0004-6361/202141249},
  \href {https://ui.adsabs.harvard.edu/abs/2021A&A...652A..86C} {652, A86}

\bibitem[\protect\citeauthoryear{{Carretta}, {Gratton}, {Cohen}, {Beers}  \&
  {Christlieb}}{{Carretta} et~al.}{2002}]{Carretta2002}
{Carretta} E.,  {Gratton} R.,  {Cohen} J.~G.,  {Beers} T.~C.,   {Christlieb}
  N.,  2002, \mn@doi [\aj] {10.1086/340955}, \href
  {https://ui.adsabs.harvard.edu/abs/2002AJ....124..481C} {124, 481}

\bibitem[\protect\citeauthoryear{{Casagrande} \& {VandenBerg}}{{Casagrande} \&
  {VandenBerg}}{2018a}]{Casagrande2018BC1}
{Casagrande} L.,  {VandenBerg} D.~A.,  2018a, \mn@doi [\mnras]
  {10.1093/mnras/sty149}, \href
  {https://ui.adsabs.harvard.edu/abs/2018MNRAS.475.5023C} {475, 5023}

\bibitem[\protect\citeauthoryear{{Casagrande} \& {VandenBerg}}{{Casagrande} \&
  {VandenBerg}}{2018b}]{Casagrande2018BC2}
{Casagrande} L.,  {VandenBerg} D.~A.,  2018b, \mn@doi [\mnras]
  {10.1093/mnrasl/sly104}, \href
  {https://ui.adsabs.harvard.edu/abs/2018MNRAS.479L.102C} {479, L102}

\bibitem[\protect\citeauthoryear{{Casey}}{{Casey}}{2014}]{casey14}
{Casey} A.~R.,  2014, PhD thesis, Australian National University, Canberra

\bibitem[\protect\citeauthoryear{{Casey} \& {Schlaufman}}{{Casey} \&
  {Schlaufman}}{2015}]{Casey2015}
{Casey} A.~R.,  {Schlaufman} K.~C.,  2015, \mn@doi [\apj]
  {10.1088/0004-637X/809/2/110}, \href
  {https://ui.adsabs.harvard.edu/abs/2015ApJ...809..110C} {809, 110}

\bibitem[\protect\citeauthoryear{{Castelli} \& {Kurucz}}{{Castelli} \&
  {Kurucz}}{2004}]{Castelli2004}
{Castelli} F.,  {Kurucz} R.~L.,  2004, \mn@doi [ArXiv Astrophysics e-prints]
  {10.48550/arXiv.astro-ph/0405087}, \href
  {http://adsabs.harvard.edu/abs/2004astro.ph..5087C} {}

\bibitem[\protect\citeauthoryear{{Cayrel} et~al.,}{{Cayrel}
  et~al.}{2004}]{cayrel2004}
{Cayrel} R.,  et~al., 2004, \mn@doi [\aap] {10.1051/0004-6361:20034074}, \href
  {http://adsabs.harvard.edu/abs/2004A%26A...416.1117C} {416, 1117}

\bibitem[\protect\citeauthoryear{{Chiti}, {Frebel}, {Jerjen}, {Kim}  \&
  {Norris}}{{Chiti} et~al.}{2020}]{Chiti_SMSS_2020}
{Chiti} A.,  {Frebel} A.,  {Jerjen} H.,  {Kim} D.,   {Norris} J.~E.,  2020,
  \mn@doi [\apj] {10.3847/1538-4357/ab6d72}, \href
  {https://ui.adsabs.harvard.edu/abs/2020ApJ...891....8C} {891, 8}

\bibitem[\protect\citeauthoryear{{Chiti}, {Frebel}, {Mardini}, {Daniel}, {Ou}
  \& {Uvarova}}{{Chiti} et~al.}{2021}]{Chiti_SMSS_cat2021}
{Chiti} A.,  {Frebel} A.,  {Mardini} M.~K.,  {Daniel} T.~W.,  {Ou} X.,
  {Uvarova} A.~V.,  2021, \mn@doi [\apjs] {10.3847/1538-4365/abf73d}, \href
  {https://ui.adsabs.harvard.edu/abs/2021ApJS..254...31C} {254, 31}

\bibitem[\protect\citeauthoryear{{Christlieb} et~al.,}{{Christlieb}
  et~al.}{2002}]{Christlieb2002}
{Christlieb} N.,  et~al., 2002, \mn@doi [\nat] {10.1038/nature01142}, \href
  {https://ui.adsabs.harvard.edu/abs/2002Natur.419..904C} {419, 904}

\bibitem[\protect\citeauthoryear{{Christlieb}, {Sch{\"o}rck}, {Frebel},
  {Beers}, {Wisotzki}  \& {Reimers}}{{Christlieb} et~al.}{2008}]{old_EMP}
{Christlieb} N.,  {Sch{\"o}rck} T.,  {Frebel} A.,  {Beers} T.~C.,  {Wisotzki}
  L.,   {Reimers} D.,  2008, \mn@doi [A\&A] {10.1051/0004-6361:20078748}, 484,
  721

\bibitem[\protect\citeauthoryear{{Cohen} et~al.,}{{Cohen}
  et~al.}{2004}]{Cohen2004}
{Cohen} J.~G.,  et~al., 2004, \mn@doi [\apj] {10.1086/422576}, \href
  {https://ui.adsabs.harvard.edu/abs/2004ApJ...612.1107C} {612, 1107}

\bibitem[\protect\citeauthoryear{{Cohen}, {Christlieb}, {Thompson},
  {McWilliam}, {Shectman}, {Reimers}, {Wisotzki}  \& {Kirby}}{{Cohen}
  et~al.}{2013}]{Cohen2013}
{Cohen} J.~G.,  {Christlieb} N.,  {Thompson} I.,  {McWilliam} A.,  {Shectman}
  S.,  {Reimers} D.,  {Wisotzki} L.,   {Kirby} E.,  2013, \mn@doi [\apj]
  {10.1088/0004-637X/778/1/56}, \href
  {https://ui.adsabs.harvard.edu/abs/2013ApJ...778...56C} {778, 56}

\bibitem[\protect\citeauthoryear{{Da Costa} et~al.,}{{Da Costa}
  et~al.}{2019}]{SMSS_EMP}
{Da Costa} G.~S.,  et~al., 2019, \mn@doi [\mnras] {10.1093/mnras/stz2550},
  \href {https://ui.adsabs.harvard.edu/abs/2019MNRAS.489.5900D} {489, 5900}

\bibitem[\protect\citeauthoryear{{De Angeli} et~al.,}{{De Angeli}
  et~al.}{2023}]{GaiaXP3}
{De Angeli} F.,  et~al., 2023, \mn@doi [\aap] {10.1051/0004-6361/202243680},
  \href {https://ui.adsabs.harvard.edu/abs/2023A&A...674A...2D} {674, A2}

\bibitem[\protect\citeauthoryear{{Depagne}, {Hill}, {Christlieb}  \&
  {Primas}}{{Depagne} et~al.}{2000}]{Depagne2000}
{Depagne} E.,  {Hill} V.,  {Christlieb} N.,   {Primas} F.,  2000, \aap, \href
  {https://ui.adsabs.harvard.edu/abs/2000A&A...364L...6D} {364, L6}

\bibitem[\protect\citeauthoryear{{Ezzeddine} et~al.,}{{Ezzeddine}
  et~al.}{2019}]{ezzeddine19}
{Ezzeddine} R.,  et~al., 2019, \mn@doi [\apj] {10.3847/1538-4357/ab14e7}, \href
  {https://ui.adsabs.harvard.edu/abs/2019ApJ...876...97E} {876, 97}

\bibitem[\protect\citeauthoryear{{Ezzeddine} et~al.,}{{Ezzeddine}
  et~al.}{2020}]{Ezzeddine_RPA_2020}
{Ezzeddine} R.,  et~al., 2020, \mn@doi [\apj] {10.3847/1538-4357/ab9d1a}, \href
  {https://ui.adsabs.harvard.edu/abs/2020ApJ...898..150E} {898, 150}

\bibitem[\protect\citeauthoryear{{For} \& {Sneden}}{{For} \&
  {Sneden}}{2010}]{For2010}
{For} B.-Q.,  {Sneden} C.,  2010, \mn@doi [\aj] {10.1088/0004-6256/140/6/1694},
  \href {https://ui.adsabs.harvard.edu/abs/2010AJ....140.1694F} {140, 1694}

\bibitem[\protect\citeauthoryear{{Francois} et~al.,}{{Francois}
  et~al.}{2007}]{francois07}
{Francois} P.,  et~al., 2007, \mn@doi [A\&A] {10.1051/0004-6361:20077706}, 476,
  935

\bibitem[\protect\citeauthoryear{{Frebel}}{{Frebel}}{2018}]{Frebel2018}
{Frebel} A.,  2018, \mn@doi [Annual Review of Nuclear and Particle Science]
  {10.1146/annurev-nucl-101917-021141}, \href
  {https://ui.adsabs.harvard.edu/abs/2018ARNPS..68..237F} {68, 237}

\bibitem[\protect\citeauthoryear{{Frebel} \& {Norris}}{{Frebel} \&
  {Norris}}{2013}]{fn13}
{Frebel} A.,  {Norris} J.~E.,  2013, {Metal-Poor Stars and the Chemical
  Enrichment of the Universe in Planets, Stars and Stellar Systems.~Volume~5:
  Galactic Structure and Stellar Populations, eds. {Oswalt}, T.~D. and
  {Gilmore}, G.}.
pp. 55-114, Dordrecht: Springer Science+Business Media,
  \mn@doi{10.1007/978-94-007-5612-0_3}

\bibitem[\protect\citeauthoryear{{Frebel} \& {Norris}}{{Frebel} \&
  {Norris}}{2015}]{Frebel_Norris_2015}
{Frebel} A.,  {Norris} J.~E.,  2015, \mn@doi [\araa]
  {10.1146/annurev-astro-082214-122423}, \href
  {https://ui.adsabs.harvard.edu/abs/2015ARA&A..53..631F} {53, 631}

\bibitem[\protect\citeauthoryear{{Frebel} et~al.,}{{Frebel}
  et~al.}{2005}]{HE1327_Nature}
{Frebel} A.,  et~al., 2005, \mn@doi [\nat] {10.1038/nature03455}, \href
  {http://adsabs.harvard.edu/abs/2005Natur.434..871F} {434, 871}

\bibitem[\protect\citeauthoryear{{Frebel} et~al.,}{{Frebel}
  et~al.}{2006}]{frebel_EMP}
{Frebel} A.,  et~al., 2006, \mn@doi [ApJ] {10.1086/508506}, 652, 1585

\bibitem[\protect\citeauthoryear{{Frebel}, {Johnson}  \& {Bromm}}{{Frebel}
  et~al.}{2007}]{dtrans}
{Frebel} A.,  {Johnson} J.~L.,   {Bromm} V.,  2007, \mn@doi [MNRAS]
  {10.1111/j.1745-3933.2007.00344.x}, 380, L40

\bibitem[\protect\citeauthoryear{{Frebel}, {Collet}, {Eriksson}, {Christlieb}
  \& {Aoki}}{{Frebel} et~al.}{2008}]{frebel08}
{Frebel} A.,  {Collet} R.,  {Eriksson} K.,  {Christlieb} N.,   {Aoki} W.,
  2008, \mn@doi [\apj] {10.1086/590327}, \href
  {http://adsabs.harvard.edu/abs/2008ApJ...684..588F} {684, 588}

\bibitem[\protect\citeauthoryear{{Frebel}, {Simon}, {Geha}  \&
  {Willman}}{{Frebel} et~al.}{2010}]{Frebel2010}
{Frebel} A.,  {Simon} J.~D.,  {Geha} M.,   {Willman} B.,  2010, \mn@doi [\apj]
  {10.1088/0004-637X/708/1/560}, \href
  {https://ui.adsabs.harvard.edu/abs/2010ApJ...708..560F} {708, 560}

\bibitem[\protect\citeauthoryear{{Frebel}, {Casey}, {Jacobson}  \&
  {Yu}}{{Frebel} et~al.}{2013}]{frebel13}
{Frebel} A.,  {Casey} A.~R.,  {Jacobson} H.~R.,   {Yu} Q.,  2013, \mn@doi [ApJ]
  {10.1088/0004-637X/769/1/57}, 769, 57

\bibitem[\protect\citeauthoryear{{Frebel}, {Simon}  \& {Kirby}}{{Frebel}
  et~al.}{2014}]{frebel14}
{Frebel} A.,  {Simon} J.~D.,   {Kirby} E.~N.,  2014, \mn@doi [ApJ]
  {10.1088/0004-637X/786/1/74}, \href
  {http://adsabs.harvard.edu/abs/2014ApJ...786...74F} {786, 74}

\bibitem[\protect\citeauthoryear{{Frebel}, {Chiti}, {Ji}, {Jacobson}  \&
  {Placco}}{{Frebel} et~al.}{2015}]{frebel15b}
{Frebel} A.,  {Chiti} A.,  {Ji} A.,  {Jacobson} H.~R.,   {Placco} V.,  2015,
  \mn@doi [ApJL] {10.1088/2041-8205/810/2/L27}, \href
  {http://adsabs.harvard.edu/abs/2015ApJ...810L..27F} {810, L27}

\bibitem[\protect\citeauthoryear{{Frebel}, {Ji}, {Ezzeddine}, {Hansen},
  {Chiti}, {Thompson}  \& {Merle}}{{Frebel} et~al.}{2019}]{Frebel2019}
{Frebel} A.,  {Ji} A.~P.,  {Ezzeddine} R.,  {Hansen} T.~T.,  {Chiti} A.,
  {Thompson} I.~B.,   {Merle} T.,  2019, \mn@doi [\apj]
  {10.3847/1538-4357/aae848}, \href
  {https://ui.adsabs.harvard.edu/abs/2019ApJ...871..146F} {871, 146}

\bibitem[\protect\citeauthoryear{{Frischknecht} et~al.,}{{Frischknecht}
  et~al.}{2016}]{Frischknecht_Sr_Ba}
{Frischknecht} U.,  et~al., 2016, \mn@doi [\mnras] {10.1093/mnras/stv2723},
  \href {https://ui.adsabs.harvard.edu/abs/2016MNRAS.456.1803F} {456, 1803}

\bibitem[\protect\citeauthoryear{{Gaia Collaboration} et~al.,}{{Gaia
  Collaboration} et~al.}{2016}]{Gaia_the_mission}
{Gaia Collaboration} et~al., 2016, \mn@doi [\aap]
  {10.1051/0004-6361/201629272}, \href
  {https://ui.adsabs.harvard.edu/abs/2016A&A...595A...1G} {595, A1}

\bibitem[\protect\citeauthoryear{{Gaia Collaboration} et~al.,}{{Gaia
  Collaboration} et~al.}{2023}]{Gaia_DR3}
{Gaia Collaboration} et~al., 2023, \mn@doi [\aap]
  {10.1051/0004-6361/202243940}, \href
  {https://ui.adsabs.harvard.edu/abs/2023A&A...674A...1G} {674, A1}

\bibitem[\protect\citeauthoryear{{Gallagher}, {Ryan}, {Garc{\'\i}a P{\'e}rez}
  \& {Aoki}}{{Gallagher} et~al.}{2010}]{Gallagher2010}
{Gallagher} A.~J.,  {Ryan} S.~G.,  {Garc{\'\i}a P{\'e}rez} A.~E.,   {Aoki} W.,
  2010, \mn@doi [\aap] {10.1051/0004-6361/201014970}, \href
  {https://ui.adsabs.harvard.edu/abs/2010A&A...523A..24G} {523, A24}

\bibitem[\protect\citeauthoryear{{Gardner} et~al.,}{{Gardner}
  et~al.}{2006}]{JWST_telescope}
{Gardner} J.~P.,  et~al., 2006, \mn@doi [\ssr] {10.1007/s11214-006-8315-7},
  \href {https://ui.adsabs.harvard.edu/abs/2006SSRv..123..485G} {123, 485}

\bibitem[\protect\citeauthoryear{{Hansen}, {Bergemann}, {Cescutti},
  {Fran{\c{c}}ois}, {Arcones}, {Karakas}, {Lind}  \& {Chiappini}}{{Hansen}
  et~al.}{2013}]{Hansen2013}
{Hansen} C.~J.,  {Bergemann} M.,  {Cescutti} G.,  {Fran{\c{c}}ois} P.,
  {Arcones} A.,  {Karakas} A.~I.,  {Lind} K.,   {Chiappini} C.,  2013, \mn@doi
  [\aap] {10.1051/0004-6361/201220584}, \href
  {https://ui.adsabs.harvard.edu/abs/2013A&A...551A..57H} {551, A57}

\bibitem[\protect\citeauthoryear{{Hansen} et~al.,}{{Hansen}
  et~al.}{2015}]{Hansen2015}
{Hansen} T.,  et~al., 2015, \mn@doi [\apj] {10.1088/0004-637X/807/2/173}, \href
  {https://ui.adsabs.harvard.edu/abs/2015ApJ...807..173H} {807, 173}

\bibitem[\protect\citeauthoryear{{Hansen}, {El-Souri}, {Monaco}, {Villanova},
  {Bonifacio}, {Caffau}  \& {Sbordone}}{{Hansen} et~al.}{2018}]{Hansen2018_Sgr}
{Hansen} C.~J.,  {El-Souri} M.,  {Monaco} L.,  {Villanova} S.,  {Bonifacio} P.,
   {Caffau} E.,   {Sbordone} L.,  2018, \mn@doi [\apj]
  {10.3847/1538-4357/aa978f}, \href
  {https://ui.adsabs.harvard.edu/abs/2018ApJ...855...83H} {855, 83}

\bibitem[\protect\citeauthoryear{{Hartwig} et~al.,}{{Hartwig}
  et~al.}{2018}]{Tilman_mono}
{Hartwig} T.,  et~al., 2018, \mn@doi [\mnras] {10.1093/mnras/sty1176}, \href
  {https://ui.adsabs.harvard.edu/abs/2018MNRAS.478.1795H} {478, 1795}

\bibitem[\protect\citeauthoryear{{Heger} \& {Woosley}}{{Heger} \&
  {Woosley}}{2010}]{heger_woosley10}
{Heger} A.,  {Woosley} S.~E.,  2010, \mn@doi [ApJ]
  {10.1088/0004-637X/724/1/341}, \href
  {http://adsabs.harvard.edu/abs/2010ApJ...724..341H} {724, 341}

\bibitem[\protect\citeauthoryear{{Hollek}, {Frebel}, {Roederer}, {Sneden},
  {Shetrone}, {Beers}, {Kang}  \& {Thom}}{{Hollek} et~al.}{2011}]{Hollek2011}
{Hollek} J.~K.,  {Frebel} A.,  {Roederer} I.~U.,  {Sneden} C.,  {Shetrone} M.,
  {Beers} T.~C.,  {Kang} S.-j.,   {Thom} C.,  2011, \mn@doi [\apj]
  {10.1088/0004-637X/742/1/54}, \href
  {https://ui.adsabs.harvard.edu/abs/2011ApJ...742...54H} {742, 54}

\bibitem[\protect\citeauthoryear{{Honda}, {Aoki}, {Beers}  \&
  {Takada-Hidai}}{{Honda} et~al.}{2011}]{Honda2011}
{Honda} S.,  {Aoki} W.,  {Beers} T.~C.,   {Takada-Hidai} M.,  2011, \mn@doi
  [\apj] {10.1088/0004-637X/730/2/77}, \href
  {https://ui.adsabs.harvard.edu/abs/2011ApJ...730...77H} {730, 77}

\bibitem[\protect\citeauthoryear{{Jacobson} et~al.,}{{Jacobson}
  et~al.}{2015}]{Jacobson2015}
{Jacobson} H.~R.,  et~al., 2015, \mn@doi [\apj] {10.1088/0004-637X/807/2/171},
  \href {https://ui.adsabs.harvard.edu/abs/2015ApJ...807..171J} {807, 171}

\bibitem[\protect\citeauthoryear{Keeping}{Keeping}{1962}]{Keeping62}
Keeping E.~S.,  1962, Introduction to statistical inference..
Princeton, N.J., Van Nostrand [1962], \url
  {https://archive.org/details/introductiontost00keep/page/52/mode/2up}

\bibitem[\protect\citeauthoryear{{Keller} et~al.,}{{Keller}
  et~al.}{2014}]{Keller2014}
{Keller} S.~C.,  et~al., 2014, \mn@doi [\nat] {10.1038/nature12990}, \href
  {https://ui.adsabs.harvard.edu/abs/2014Natur.506..463K} {506, 463}

\bibitem[\protect\citeauthoryear{{Kelson}}{{Kelson}}{2003}]{kelson03}
{Kelson} D.~D.,  2003, \mn@doi [\pasp] {10.1086/375502}, 115, 688

\bibitem[\protect\citeauthoryear{{Kunder} et~al.,}{{Kunder}
  et~al.}{2017}]{RAVE_5th}
{Kunder} A.,  et~al., 2017, \mn@doi [\aj] {10.3847/1538-3881/153/2/75}, \href
  {https://ui.adsabs.harvard.edu/abs/2017AJ....153...75K} {153, 75}

\bibitem[\protect\citeauthoryear{{Lai}, {Bolte}, {Johnson}  \&
  {Lucatello}}{{Lai} et~al.}{2004}]{Lai2004}
{Lai} D.~K.,  {Bolte} M.,  {Johnson} J.~A.,   {Lucatello} S.,  2004, \mn@doi
  [\aj] {10.1086/424864}, \href
  {https://ui.adsabs.harvard.edu/abs/2004AJ....128.2402L} {128, 2402}

\bibitem[\protect\citeauthoryear{{Lai}, {Bolte}, {Johnson}, {Lucatello},
  {Heger}  \& {Woosley}}{{Lai} et~al.}{2008}]{Lai2008}
{Lai} D.~K.,  {Bolte} M.,  {Johnson} J.~A.,  {Lucatello} S.,  {Heger} A.,
  {Woosley} S.~E.,  2008, \mn@doi [ApJ] {10.1086/588811}, 681, 1524

\bibitem[\protect\citeauthoryear{{Magg} et~al.,}{{Magg}
  et~al.}{2020}]{Magg2020_fits}
{Magg} M.,  et~al., 2020, \mn@doi [\mnras] {10.1093/mnras/staa2624}, \href
  {https://ui.adsabs.harvard.edu/abs/2020MNRAS.498.3703M} {498, 3703}

\bibitem[\protect\citeauthoryear{Mardini et~al.,}{Mardini
  et~al.}{2019a}]{Mardini_2019a}
Mardini M.~K.,  et~al., 2019a, \mn@doi [The Astrophysical Journal]
  {10.3847/1538-4357/ab0fa2}, 875, 89

\bibitem[\protect\citeauthoryear{Mardini, Placco, Taani, Li  \& Zhao}{Mardini
  et~al.}{2019b}]{Mardini_2019b}
Mardini M.~K.,  Placco V.~M.,  Taani A.,  Li H.,   Zhao G.,  2019b, \mn@doi
  [The Astrophysical Journal] {10.3847/1538-4357/ab3047}, 882, 27

\bibitem[\protect\citeauthoryear{{Mardini}, {Ershiadat}, {Al-Wardat}, {Taani},
  {{\"O}zdemir}, {Al-Naimiy}  \& {Khasawneh}}{{Mardini}
  et~al.}{2019c}]{Mardini2019c}
{Mardini} M.~K.,  {Ershiadat} N.,  {Al-Wardat} M.~A.,  {Taani} A.~A.,
  {{\"O}zdemir} S.,  {Al-Naimiy} H.,   {Khasawneh} A.,  2019c, in Journal of
  Physics Conference Series. p. 012024 (\mn@eprint {arXiv} {1904.09608}),
  \mn@doi{10.1088/1742-6596/1258/1/012024}

\bibitem[\protect\citeauthoryear{Mardini et~al.,}{Mardini
  et~al.}{2020}]{Mardini_2020}
Mardini M.~K.,  et~al., 2020, \mn@doi [The Astrophysical Journal]
  {10.3847/1538-4357/abbc13}, 903, 88

\bibitem[\protect\citeauthoryear{{Mardini} et~al.,}{{Mardini}
  et~al.}{2022a}]{Mardini2022b}
{Mardini} M.~K.,  et~al., 2022a, \mn@doi [\mnras] {10.1093/mnras/stac2783},
  \href {https://ui.adsabs.harvard.edu/abs/2022MNRAS.517.3993M} {517, 3993}

\bibitem[\protect\citeauthoryear{{Mardini}, {Frebel}, {Chiti}, {Meiron},
  {Brauer}  \& {Ou}}{{Mardini} et~al.}{2022b}]{Mardini2022}
{Mardini} M.~K.,  {Frebel} A.,  {Chiti} A.,  {Meiron} Y.,  {Brauer} K.~V.,
  {Ou} X.,  2022b, \mn@doi [\apj] {10.3847/1538-4357/ac8102}, \href
  {https://ui.adsabs.harvard.edu/abs/2022ApJ...936...78M} {936, 78}

\bibitem[\protect\citeauthoryear{{Mardini}, {Frebel}, {Betre}, {Jacobson},
  {Norris}  \& {Christlieb}}{{Mardini} et~al.}{2023}]{Mardini23_dwarf}
{Mardini} M.~K.,  {Frebel} A.,  {Betre} L.,  {Jacobson} H.,  {Norris} J.~E.,
  {Christlieb} N.,  2023, \mn@doi [arXiv e-prints] {10.48550/arXiv.2305.05363},
  \href {https://ui.adsabs.harvard.edu/abs/2023arXiv230505363M} {p.
  arXiv:2305.05363}

\bibitem[\protect\citeauthoryear{{Mashonkina}, {Jablonka}, {Sitnova},
  {Pakhomov}  \& {North}}{{Mashonkina} et~al.}{2017}]{Mashonkina2017}
{Mashonkina} L.,  {Jablonka} P.,  {Sitnova} T.,  {Pakhomov} Y.,   {North} P.,
  2017, \mn@doi [\aap] {10.1051/0004-6361/201731582}, \href
  {https://ui.adsabs.harvard.edu/abs/2017A&A...608A..89M} {608, A89}

\bibitem[\protect\citeauthoryear{{Masseron} et~al.,}{{Masseron}
  et~al.}{2006}]{Masseron2006}
{Masseron} T.,  et~al., 2006, \mn@doi [A\&A] {10.1051/0004-6361:20064802}, 455,
  1059

\bibitem[\protect\citeauthoryear{{Montegriffo} et~al.,}{{Montegriffo}
  et~al.}{2023}]{GaiaXP2}
{Montegriffo} P.,  et~al., 2023, \mn@doi [\aap] {10.1051/0004-6361/202243880},
  \href {https://ui.adsabs.harvard.edu/abs/2023A&A...674A...3M} {674, A3}

\bibitem[\protect\citeauthoryear{{Mucciarelli}, {Bellazzini}  \&
  {Massari}}{{Mucciarelli} et~al.}{2021}]{Mucciarelli2021}
{Mucciarelli} A.,  {Bellazzini} M.,   {Massari} D.,  2021, \mn@doi [\aap]
  {10.1051/0004-6361/202140979}, \href
  {https://ui.adsabs.harvard.edu/abs/2021A&A...653A..90M} {653, A90}

\bibitem[\protect\citeauthoryear{{Norris}, {Ryan}  \& {Beers}}{{Norris}
  et~al.}{2001}]{Norris2001}
{Norris} J.~E.,  {Ryan} S.~G.,   {Beers} T.~C.,  2001, \mn@doi [\apj]
  {10.1086/323429}, \href
  {https://ui.adsabs.harvard.edu/abs/2001ApJ...561.1034N} {561, 1034}

\bibitem[\protect\citeauthoryear{{Norris}, {Christlieb}, {Korn}, {Eriksson},
  {Bessell}, {Beers}, {Wisotzki}  \& {Reimers}}{{Norris}
  et~al.}{2007}]{Norris2007}
{Norris} J.~E.,  {Christlieb} N.,  {Korn} A.~J.,  {Eriksson} K.,  {Bessell}
  M.~S.,  {Beers} T.~C.,  {Wisotzki} L.,   {Reimers} D.,  2007, \mn@doi [\apj]
  {10.1086/521919}, \href
  {https://ui.adsabs.harvard.edu/abs/2007ApJ...670..774N} {670, 774}

\bibitem[\protect\citeauthoryear{{Placco}, {Frebel}, {Beers}, {Christlieb},
  {Lee}, {Kennedy}, {Rossi}  \& {Santucci}}{{Placco}
  et~al.}{2014a}]{placco2014a}
{Placco} V.~M.,  {Frebel} A.,  {Beers} T.~C.,  {Christlieb} N.,  {Lee} Y.~S.,
  {Kennedy} C.~R.,  {Rossi} S.,   {Santucci} R.~M.,  2014a, \mn@doi [\apj]
  {10.1088/0004-637X/781/1/40}, \href
  {https://ui.adsabs.harvard.edu/abs/2014ApJ...781...40P} {781, 40}

\bibitem[\protect\citeauthoryear{{Placco}, {Frebel}, {Beers}  \&
  {Stancliffe}}{{Placco} et~al.}{2014b}]{placco14_carbon}
{Placco} V.~M.,  {Frebel} A.,  {Beers} T.~C.,   {Stancliffe} R.~J.,  2014b,
  \mn@doi [ApJ] {10.1088/0004-637X/797/1/21}, 797, 21

\bibitem[\protect\citeauthoryear{{Placco} et~al.,}{{Placco}
  et~al.}{2016}]{placco16}
{Placco} V.~M.,  et~al., 2016, \mn@doi [\apj] {10.3847/0004-637X/833/1/21},
  \href {http://adsabs.harvard.edu/abs/2016ApJ...833...21P} {833, 21}

\bibitem[\protect\citeauthoryear{{Placco} et~al.,}{{Placco}
  et~al.}{2020}]{Placco2020}
{Placco} V.~M.,  et~al., 2020, \mn@doi [\apj] {10.3847/1538-4357/ab99c6}, \href
  {https://ui.adsabs.harvard.edu/abs/2020ApJ...897...78P} {897, 78}

\bibitem[\protect\citeauthoryear{{Placco}, {Sneden}, {Roederer}, {Lawler}, {Den
  Hartog}, {Hejazi}, {Maas}  \& {Bernath}}{{Placco}
  et~al.}{2021a}]{Placco2021_linemake}
{Placco} V.~M.,  {Sneden} C.,  {Roederer} I.~U.,  {Lawler} J.~E.,  {Den Hartog}
  E.~A.,  {Hejazi} N.,  {Maas} Z.,   {Bernath} P.,  2021a, \mn@doi [Research
  Notes of the American Astronomical Society] {10.3847/2515-5172/abf651}, \href
  {https://ui.adsabs.harvard.edu/abs/2021RNAAS...5...92P} {5, 92}

\bibitem[\protect\citeauthoryear{{Placco} et~al.,}{{Placco}
  et~al.}{2021b}]{Placco2021}
{Placco} V.~M.,  et~al., 2021b, \mn@doi [\apjl] {10.3847/2041-8213/abf93d},
  \href {https://ui.adsabs.harvard.edu/abs/2021ApJ...912L..32P} {912, L32}

\bibitem[\protect\citeauthoryear{{Placco}, {Almeida-Fernandes}, {Arentsen},
  {Lee}, {Schoenell}, {Ribeiro}  \& {Kanaan}}{{Placco}
  et~al.}{2022}]{Placco2022}
{Placco} V.~M.,  {Almeida-Fernandes} F.,  {Arentsen} A.,  {Lee} Y.~S.,
  {Schoenell} W.,  {Ribeiro} T.,   {Kanaan} A.,  2022, \mn@doi [\apjs]
  {10.3847/1538-4365/ac7ab0}, \href
  {https://ui.adsabs.harvard.edu/abs/2022ApJS..262....8P} {262, 8}

\bibitem[\protect\citeauthoryear{{Placco} et~al.,}{{Placco}
  et~al.}{2023}]{placco2023}
{Placco} V.~M.,  et~al., 2023, \mn@doi [\apj] {10.3847/1538-4357/ad077e}, \href
  {https://ui.adsabs.harvard.edu/abs/2023ApJ...959...60P} {959, 60}

\bibitem[\protect\citeauthoryear{{Plez} \& {Cohen}}{{Plez} \&
  {Cohen}}{2005}]{Plez2005}
{Plez} B.,  {Cohen} J.~G.,  2005, \mn@doi [\aap] {10.1051/0004-6361:20042082},
  \href {https://ui.adsabs.harvard.edu/abs/2005A&A...434.1117P} {434, 1117}

\bibitem[\protect\citeauthoryear{{Pruet}, {Hoffman}, {Woosley}, {Janka}  \&
  {Buras}}{{Pruet} et~al.}{2006}]{Sr_from_CCSN2}
{Pruet} J.,  {Hoffman} R.~D.,  {Woosley} S.~E.,  {Janka} H.~T.,   {Buras} R.,
  2006, \mn@doi [\apj] {10.1086/503891}, \href
  {https://ui.adsabs.harvard.edu/abs/2006ApJ...644.1028P} {644, 1028}

\bibitem[\protect\citeauthoryear{{Rich} \& {Boesgaard}}{{Rich} \&
  {Boesgaard}}{2009}]{Rich2009}
{Rich} J.~A.,  {Boesgaard} A.~M.,  2009, \mn@doi [\apj]
  {10.1088/0004-637X/701/2/1519}, \href
  {https://ui.adsabs.harvard.edu/abs/2009ApJ...701.1519R} {701, 1519}

\bibitem[\protect\citeauthoryear{{Rix} et~al.,}{{Rix} et~al.}{2022}]{Rix2022}
{Rix} H.-W.,  et~al., 2022, \mn@doi [\apj] {10.3847/1538-4357/ac9e01}, \href
  {https://ui.adsabs.harvard.edu/abs/2022ApJ...941...45R} {941, 45}

\bibitem[\protect\citeauthoryear{{Roederer}, {Preston}, {Thompson}, {Shectman},
  {Sneden}, {Burley}  \& {Kelson}}{{Roederer} et~al.}{2014}]{roederer14c}
{Roederer} I.~U.,  {Preston} G.~W.,  {Thompson} I.~B.,  {Shectman} S.~A.,
  {Sneden} C.,  {Burley} G.~S.,   {Kelson} D.~D.,  2014, \mn@doi [\aj]
  {10.1088/0004-6256/147/6/136}, \href
  {http://adsabs.harvard.edu/abs/2014AJ....147..136R} {147, 136}

\bibitem[\protect\citeauthoryear{{Rossi}, {Salvadori}, {Sk{\'u}lad{\'o}ttir}
  \& {Vanni}}{{Rossi} et~al.}{2023}]{Rossi_CEMPno_pop3}
{Rossi} M.,  {Salvadori} S.,  {Sk{\'u}lad{\'o}ttir} {\'A}.,   {Vanni} I.,
  2023, \mn@doi [\mnras] {10.1093/mnrasl/slad029}, \href
  {https://ui.adsabs.harvard.edu/abs/2023MNRAS.522L...1R} {522, L1}

\bibitem[\protect\citeauthoryear{{Ryan}, {Norris}  \& {Bessell}}{{Ryan}
  et~al.}{1991}]{Ryan1991}
{Ryan} S.~G.,  {Norris} J.~E.,   {Bessell} M.~S.,  1991, \mn@doi [\aj]
  {10.1086/115878}, \href
  {https://ui.adsabs.harvard.edu/abs/1991AJ....102..303R} {102, 303}

\bibitem[\protect\citeauthoryear{{Ryan}, {Norris}  \& {Beers}}{{Ryan}
  et~al.}{1996}]{Ryan1996}
{Ryan} S.~G.,  {Norris} J.~E.,   {Beers} T.~C.,  1996, \mn@doi [\apj]
  {10.1086/177967}, \href
  {https://ui.adsabs.harvard.edu/abs/1996ApJ...471..254R} {471, 254}

\bibitem[\protect\citeauthoryear{{Ryan}, {Norris}  \& {Beers}}{{Ryan}
  et~al.}{1999}]{Ryan1999}
{Ryan} S.~G.,  {Norris} J.~E.,   {Beers} T.~C.,  1999, \mn@doi [\apj]
  {10.1086/307769}, \href
  {https://ui.adsabs.harvard.edu/abs/1999ApJ...523..654R} {523, 654}

\bibitem[\protect\citeauthoryear{{Sivarani} et~al.,}{{Sivarani}
  et~al.}{2006}]{Sivarani2006}
{Sivarani} T.,  et~al., 2006, \mn@doi [\aap] {10.1051/0004-6361:20065440},
  \href {https://ui.adsabs.harvard.edu/abs/2006A&A...459..125S} {459, 125}

\bibitem[\protect\citeauthoryear{{Sneden}}{{Sneden}}{1973}]{moog}
{Sneden} C.~A.,  1973, PhD thesis, University of Texas, Austin

\bibitem[\protect\citeauthoryear{{Spite}, {Spite}, {Cayrel}, {Hill},
  {Nordstr{\"o}m}, {Barbuy}, {Beers}  \& {Nissen}}{{Spite}
  et~al.}{1999}]{Spite1999}
{Spite} M.,  {Spite} F.,  {Cayrel} R.,  {Hill} V.,  {Nordstr{\"o}m} B.,
  {Barbuy} B.,  {Beers} T.,   {Nissen} P.~E.,  1999, \mn@doi [\apss]
  {10.1023/A:1002170401475}, \href
  {https://ui.adsabs.harvard.edu/abs/1999Ap&SS.265..141S} {265, 141}

\bibitem[\protect\citeauthoryear{{Spite}, {Depagne}, {Nordstr{\"o}m}, {Hill},
  {Cayrel}, {Spite}  \& {Beers}}{{Spite} et~al.}{2000}]{Spite2000}
{Spite} M.,  {Depagne} E.,  {Nordstr{\"o}m} B.,  {Hill} V.,  {Cayrel} R.,
  {Spite} F.,   {Beers} T.~C.,  2000, \aap, \href
  {https://ui.adsabs.harvard.edu/abs/2000A&A...360.1077S} {360, 1077}

\bibitem[\protect\citeauthoryear{{Spite}, {Spite}, {Bonifacio}, {Caffau},
  {Fran{\c{c}}ois}  \& {Sbordone}}{{Spite} et~al.}{2014}]{Spite2014}
{Spite} M.,  {Spite} F.,  {Bonifacio} P.,  {Caffau} E.,  {Fran{\c{c}}ois} P.,
  {Sbordone} L.,  2014, \mn@doi [\aap] {10.1051/0004-6361/201423658}, \href
  {https://ui.adsabs.harvard.edu/abs/2014A&A...571A..40S} {571, A40}

\bibitem[\protect\citeauthoryear{{Starkenburg} et~al.,}{{Starkenburg}
  et~al.}{2017}]{pristine_EMP}
{Starkenburg} E.,  et~al., 2017, \mn@doi [\mnras] {10.1093/mnras/stx1068},
  \href {https://ui.adsabs.harvard.edu/abs/2017MNRAS.471.2587S} {471, 2587}

\bibitem[\protect\citeauthoryear{{Steinmetz} et~al.,}{{Steinmetz}
  et~al.}{2020}]{RAVE_6th}
{Steinmetz} M.,  et~al., 2020, \mn@doi [\aj] {10.3847/1538-3881/ab9ab8}, \href
  {https://ui.adsabs.harvard.edu/abs/2020AJ....160...83S} {160, 83}

\bibitem[\protect\citeauthoryear{{Umeda} \& {Nomoto}}{{Umeda} \&
  {Nomoto}}{2005}]{umeda&nomoto05}
{Umeda} H.,  {Nomoto} K.,  2005, \mn@doi [\apj] {10.1086/426097}, \href
  {http://adsabs.harvard.edu/abs/2005ApJ...619..427U} {619, 427}

\bibitem[\protect\citeauthoryear{{Wanajo}, {Janka}  \& {M{\"u}ller}}{{Wanajo}
  et~al.}{2011}]{Wanajo2011}
{Wanajo} S.,  {Janka} H.-T.,   {M{\"u}ller} B.,  2011, \mn@doi [\apjl]
  {10.1088/2041-8205/726/2/L15}, \href
  {https://ui.adsabs.harvard.edu/abs/2011ApJ...726L..15W} {726, L15}

\bibitem[\protect\citeauthoryear{{Watson} et~al.,}{{Watson}
  et~al.}{2019}]{Watson_Sr_detection}
{Watson} D.,  et~al., 2019, \mn@doi [\nat] {10.1038/s41586-019-1676-3}, \href
  {https://ui.adsabs.harvard.edu/abs/2019Natur.574..497W} {574, 497}

\bibitem[\protect\citeauthoryear{{Whitten} et~al.,}{{Whitten}
  et~al.}{2021}]{Whitten2021}
{Whitten} D.~D.,  et~al., 2021, \mn@doi [\apj] {10.3847/1538-4357/abee7e},
  \href {https://ui.adsabs.harvard.edu/abs/2021ApJ...912..147W} {912, 147}

\bibitem[\protect\citeauthoryear{{Yao}, {Ji}, {Koposov}  \& {Limberg}}{{Yao}
  et~al.}{2023}]{Yao2023}
{Yao} Y.,  {Ji} A.~P.,  {Koposov} S.~E.,   {Limberg} G.,  2023, \mn@doi [arXiv
  e-prints] {10.48550/arXiv.2303.17676}, \href
  {https://ui.adsabs.harvard.edu/abs/2023arXiv230317676Y} {p. arXiv:2303.17676}

\bibitem[\protect\citeauthoryear{{Yong} et~al.,}{{Yong}
  et~al.}{2013}]{Yong2013}
{Yong} D.,  et~al., 2013, \mn@doi [\apj] {10.1088/0004-637X/762/1/26}, \href
  {https://ui.adsabs.harvard.edu/abs/2013ApJ...762...26Y} {762, 26}

\bibitem[\protect\citeauthoryear{{Zhang}, {Green}  \& {Rix}}{{Zhang}
  et~al.}{2023}]{Zhang2023}
{Zhang} X.,  {Green} G.~M.,   {Rix} H.-W.,  2023, \mn@doi [\mnras]
  {10.1093/mnras/stad1941}, \href
  {https://ui.adsabs.harvard.edu/abs/2023MNRAS.tmp.1869Z} {}

\makeatother
\end{thebibliography}

\bsp	
\label{lastpage}
\end{document}